\documentclass{appolb}

\newcommand{\mra}{\mathrm{a}}
\newcommand{\mrb}{\mathrm{b}}
\newcommand{\mrc}{\mathrm{c}}
\newcommand{\mrg}{\mathrm{g}}

\newcommand{\mrm}{\mathrm{m}}
\newcommand{\mrH}{\mathrm{H}}
\newcommand{\mrO}{\mathrm{O}}
\newcommand{\mrv}{\mathrm{v}}
\newcommand{\mrM}{\mathrm{M}}
\newcommand{\mrN}{\mathrm{N}}
\newcommand{\mrD}{\mathrm{D}}
\newcommand{\mrE}{\mathrm{E}}
\newcommand{\mcO}{\mathcal{O}}
\newcommand{\mrV}{\mathrm{V}}
\newcommand{\mrZ}{\mathrm{Z}}
\newcommand{\mrW}{\mathrm{W}}
\newcommand{\mrs}{\mathrm{s}}

\newcommand{\mrA}{\mathrm{A}}
\newcommand{\mrI}{\mathrm{I}}
\newcommand{\mrJ}{\mathrm{J}}
\newcommand{\mrK}{\mathrm{K}}
\newcommand{\mrT}{\mathrm{T}}
\newcommand{\mrG}{\mathrm{G}}
\newcommand{\mrF}{\mathrm{F}}
\newcommand{\mrB}{\mathrm{B}}
\newcommand{\mrdim}{\mathrm{dim}}

\newcommand{\mrR}{\mathrm{R}}
\newcommand{\mrQ}{\mathrm{Q}}
\newcommand{\mrd}{\mathrm{d}}
\newcommand{\mrL}{\mathrm{L}}
\newcommand{\mre}{\mathrm{e}}
\newcommand{\mrf}{\mathrm{f}}

\newcommand{\mrC}{\mathrm{C}}
\newcommand{\mrS}{\mathrm{S}}
\newcommand{\eqn}[1]{Eq.(\ref{#1})}
\newcommand{\Lag}{\mathcal{L}}
\newcommand{\Ope}{\mathcal{Q}}

\newcommand{\SMEFT}{\mathrm{\scriptscriptstyle{SMEFT}}}

\newcommand{\myLO}{\mathrm{\scriptscriptstyle{LO}}}
\newcommand{\myNLO}{\mathrm{\scriptscriptstyle{NLO}}}
\newcommand{\mySM}{\mathrm{\scriptscriptstyle{SM}}}

\newcommand{\sPW}{{\scriptscriptstyle{\mrW}}}
\newcommand{\sPZ}{{\scriptscriptstyle{\mrZ}}}
\newcommand{\sPH}{{\scriptscriptstyle{\mrH}}}
\newcommand{\sPB}{{\scriptscriptstyle{\mrB}}}
\newcommand{\sPL}{{\scriptscriptstyle{\mrL}}}

\newcommand{\sPA}{{\scriptscriptstyle{\mrA}}}

\newcommand{\sPO}{{\scriptscriptstyle{\mrO}}}
\newcommand{\sPS}{{\scriptscriptstyle{\mrS}}}
\newcommand{\sPM}{{\scriptscriptstyle{\mrM}}}

\newcommand{\barnm}{\overline{\nu}_{\mu}}
\newcommand{\tDdr}{1/{\bar{\varepsilon}}}
\newcommand{\supMSB}[1]{{#1}^{\mbox{$\overline{\scriptscriptstyle MS}$}}}
\newcommand{\hsp}{\hspace{.5mm}}
\newcommand{\lpar}{\left(}                            
\newcommand{\rpar}{\right)} 
\newcommand{\stw}{\mrs_{_{\theta}}}             
\newcommand{\ctw}{\mrc_{_{\theta}}}
\newcommand{\stws}{\mrs_{_{\theta}}^2}
\newcommand{\stwc}{\mrs_{_{\theta}}^3}
\newcommand{\ctws}{\mrc_{_{\theta}}^2}
\newcommand{\ctwc}{\mrc_{_{\theta}}^3}
\newcommand{\ctwq}{\mrc_{_{\theta}}^4}
\newcommand{\spro}[2]{{#1}\cdot{#2}}
\newcommand{\PZ}{\mathrm{Z}}
\newcommand{\PA}{\mathrm{A}}
\newcommand{\PW}{\mathrm{W}}
\newcommand{\PWp}{\mathrm{W}^+}
\newcommand{\PWm}{\mathrm{W}^-}
\newcommand{\PWpm}{\mathrm{W}^{\pm}}
\newcommand{\PB}{\mathrm{B}}
\newcommand{\PX}{\mathrm{X}}
\newcommand{\PG}{\mathrm{G}}

\newcommand{\mws}{\mrM^2_{\sPW}}
\newcommand{\mzs}{\mrM^2_{\sPZ}}
\newcommand{\Pf}{\mathrm{f}}
\newcommand{\PAf}{{\overline{\mathrm{f}}}}
\newcommand{\Pl}{\mathrm{l}}
\newcommand{\PQt}{\mathrm{t}}
\newcommand{\PQb}{\mathrm{b}}
\newcommand{\PQu}{\mathrm{u}}
\newcommand{\PQd}{\mathrm{d}}
\newcommand{\PQq}{\mathrm{q}}
\newcommand{\alW}{\mra_{\Pl\scriptscriptstyle{\PW}}}
\newcommand{\alB}{\mra_{\Pl\scriptscriptstyle{\PB}}}
\newcommand{\alWB}{\mra_{\Pl\scriptscriptstyle{\PW\PB}}}
\newcommand{\alBW}{\mra_{\Pl\scriptscriptstyle{\PB\PW}}}

\newcommand{\adW}{\mra_{\PQd\scriptscriptstyle{\PW}}}

\newcommand{\adB}{\mra_{\PQd\scriptscriptstyle{\PB}}}
\newcommand{\adWB}{\mra_{\PQd\scriptscriptstyle{\PW\PB}}}
\newcommand{\adBW}{\mra_{\PQd\scriptscriptstyle{\PB\PW}}}

\newcommand{\auW}{\mra_{\PQu\scriptscriptstyle{\PW}}}

\newcommand{\auB}{\mra_{\PQu\scriptscriptstyle{\PB}}}

\newcommand{\auWB}{\mra_{\PQu\scriptscriptstyle{\PW\PB}}}
\newcommand{\auBW}{\mra_{\PQu\scriptscriptstyle{\PB\PW}}}

\newcommand{\aplo}{\mra^{(1)}_{\phi\Pl}}
\newcommand{\aplt}{\mra^{(3)}_{\phi\Pl}}
\newcommand{\apqo}{\mra^{(1)}_{\phi\PQq}}
\newcommand{\apqt}{\mra^{(3)}_{\phi\PQq}}

\newcommand{\apu}{\mra_{\phi\PQu}}
\newcommand{\apd}{\mra_{\phi\PQd}}
\newcommand{\apl}{\mra_{\phi\Pl}}
\newcommand{\auG}{\mra_{\PQu\scriptscriptstyle{\PG}}}
\newcommand{\adG}{\mra_{\PQd\scriptscriptstyle{\PG}}}
\newcommand{\afW}{\mra_{\Pf\scriptscriptstyle{\PW}}}
\newcommand{\afB}{\mra_{\Pf\scriptscriptstyle{\PB}}}
\newcommand{\Bref}[1]{Ref.~\cite{#1}}
\newcommand{\Brefs}[1]{Refs.~\cite{#1}}
\newcommand{\myeg}{e.g.\,}
\newcommand{\myie}{i.e.\,}
\newcommand{\etc}{etc.\@\,}
\newcommand{\eff}{{\mbox{\scriptsize eff}}}
\newcommand{\fin}{{\mbox{\scriptsize fin}}}
\newcommand{\DUV}{\Delta_{\mathrm{UV}}}
\usepackage[T1]{fontenc} 
\usepackage[utf8]{inputenc}
\usepackage[dvipsnames]{xcolor}

\begin{document}
\title{Veltman, renormalizability, calculability}

\author{Giampiero Passarino \\
{\small\rm{Dipartimento di Fisica Teorica, Universit\`a di Torino, Italy \\
INFN, Sezione di Torino, Italy}}}



\maketitle
\begin{abstract}
Dedicated to the memory of Prof. Veltman, one of the founding fathers of our discipline: his legacy lives on.
Many times we have to turn back and follow his footprints to find the right path.
After reviewing general aspects of high energy physics where he gave a seminal contribution we will introduce
recent developments in the standard model effective field theory, showing how the whole movement from renormalization
to predictions plays from Veltman to SMEFT.
\end{abstract}
\PACS{12.60.-i, 11.10.-z, 14.80.Bn.}

 \section{Introduction \label{Intro}}
In his talk at Higgs Hunting $2015$~\footnote{
https://indico.ijclab.in2p3.fr/event/2722/contributions/5996/} Martinus Veltman said

\noindent
\textit{Higgs particle $\,\dots\,$ What does it mean?
What does it do? It is claimed to give mass
to all other particles. What does that mean?
Can we now predict the masses of all particles?
In this talk it will be attempted to explain why the Higgs particle
is important to the theory of the Standard Model.
The importance of the Higgs construction is that it made the theory
of Yang{-}Mills fields renormalizable. Observable results can be
calculated and compared with experiment, and that has happened
in a multitude of ways in the last 40 years, up to and including the
recent discovery of the Higgs particle.}

It is clear that he meant ``strictly renormalizable'' but it is evident that the main emphasis is on
\textit{Observable results can be calculated and compared with experiment}.
In order to understand the situation before $1971$, it is interesting to observe~\cite{Veltman:2000xp}
that most of the papers on the subject came to the same conclusion `` $\dots\,$ it is concluded that all theories 
based on simple Lie groups are unrenormalizable''. 
All difficulties in the Yang{-}Mills theories disappeared in $1971$ and the theories became fully renormalizable,
that is all occurring infinities could be absorbed in the available free parameters. Theories with a Yang-Mills
structure were now renormalizable theories and a precise model for the weak interactions existed already (although it 
had received virtually no attention). The phase of ``precise calculations'' started, extending the pioneering
work of Berends, Gastmans and collaborators~\cite{Berends:1973fd,Berends:1981uq,Berends:1981rb}
to the full Standard Model~\cite{tHooft:1978jhc,Passarino:1978jh,Consoli:1979xw,Lemoine:1979pm,Antonelli:1980au}
and continued till the most recent successes~\cite{Actis:2006ra,Actis:2006rb,Actis:2006rc,Actis:2008ug}. For
a complete list of references, see \Bref{Denner:2019vbn}.
 
To summarize a long journey, we can say that the transition was from the Fermi theory to the standard model (SM); for that,
Veltman~\cite{Veltman:1968ki,Reiff:1969pq} had to convince the community that the weak interactions were some form 
of a Yang{-}Mills theory. 
As he wrote: \textit{I could not have done that without the knowledge of experimental physics that I had acquired at CERN}.
It is worth noting that the journey never contemplated the extension of the Fermi theory with the inclusion of even
higher operators; therefore, the SM represented the beyond{-}Fermi physics. 
Comparing with the present: we are now in a beyond{-}SM desert looking for alternative paths, although 
the regulative ideal of an ultimate theory remains a powerful aesthetic ingredient.

Almost equally important, and a landmark through the whole region of ``strict renormalizability'', is the work on 
quantum gravity: ``In case of gravitation interacting with scalar particles, 
divergencies in physical quantities remain''~\cite{tHooft:1974toh,Veltman:2000xp}. It is possible that at some very 
large energy scale, all nonrenormalizable interactions disappear. This seems unlikely, given the difficulty with 
gravity. It is possible that the rules change drastically. It may even be possible that there is no end, simply more 
and more scales~\cite{Georgi:1994qn}. 
\section{The standard model before LEP}
The last step in the renormalization procedure is the connection between 
renormalized quantities and physical observables. Since all quantities at this stage are
UV-free, we term it \textit{finite renormalization}.
Note that the absorption of UV divergencies into local counterterms is, to 
some extent, a trivial step (except for the problem of overlapping divergencies~\cite{Kreimer:1998iv}); 
finite renormalization, instead, requires more 
attention. For example, beyond one loop one cannot use on-shell masses but 
only complex poles for all unstable particles~\cite{Veltman:1963th,Beenakker:1996kn,Denner:2006ic,Actis:2008uh}.
The complete formulation of finite renormalization is beyond the goal of this work.
However, let us show some examples where the concept of an on-shell mass can be 
employed. Suppose that we renormalize a physical observable $\mrF$,
\begin{equation}
\mrF = \mrF_{\sPB} + g^2\,\mrF_{1\sPL}(\mrM^2) + g^4\,\mrF_{2\sPL}(\mrM^2) \hsp,
\end{equation}
where $M$ is some renormalized mass which appears at one and two loops in
$\mrF_{1\sPL}$ and $\mrF_{2\sPL}$ but does not show up in the Born term $\mrF_{\sPB}$.
In this case we can use the concept of an on-shell mass identifying
$M = M_{\sPO\sPS}$ for the two-loop term and performing a finite mass 
renormalization at one loop,
\begin{equation}
\mrM^2 = \mrM^2_{\sPO\sPS} \, \Bigl\{ 1\, +\, \frac{g^2}{16\,\pi^2}\,\Bigl[\,
\mathrm{Re}\,\Sigma^{(1)}_{\sPM}\mid_{p^2=-\mrM^2_{\sPO\sPS}} - 
\,\delta Z^{(1)}_{\sPM} \Bigr]\, \Bigr\} =
\mrM^2_{\sPO\sPS} + g^2\,\Delta \mrM^2 \hsp,
\label{stillOMS}
\end{equation}
where $M_{\sPO\sPS}$ is the on-shell mass and $\Sigma$ is extracted
from the required one-particle irreducible Green function.  It is worth noting that \eqn{stillOMS} 
is still meaningful (no dependence on gauge parameters) and will be used 
inside the one-loop result,
\begin{equation}
\mrF  = \mrF_{\sPB} + g^2\,\mrF_{1\sPL}(\mrM^2_{\sPO\sPS}) + g^4\,\Bigl[
\mrF_{2\sPL}(\mrM^2_{\sPO\sPS}) + \mrF'_{1\sPL}(\mrM^2_{\sPO\sPS})\,\Delta \mrM^2\Bigr],
\end{equation}
where
\begin{equation}
\mrF'_{1\sPL}(\mrM^2_{\sPO\sPS})\, = \,
\frac{\partial \mrF_{1\sPL}(\mrM^2)}{\partial \mrM^2}\vert_{\mrM^2=\mrM^2_{\sPO\sPS}} \hsp.
\end{equation}
If we focus on renormalization, we can safely state that all the necessary
ingredients are available. Here the crucial point is to connect a set of 
input experimental data (an input-parameter set, hereafter IPS) to the free 
parameters of the theory:
\begin{itemize}

\item[--] mass renormalization involves the calculation of 
self-energies;

\item[--] renormalization of coupling constants requires additional elements, 
which depend on the choice of the observables in the IPS.

\end{itemize}
The most-obvious selection of an IPS is based on the choice of those data 
which are known with the best experimental precision, e.g. the 
electromagnetic coupling constant, the Fermi coupling constant and the $\mathrm{Z}\,-$boson mass.

Before the advent of the LEP operations we had very few options for the IPS~\cite{Green:1980bd,Passarino:1989ey}.
To give an example, we consider a first approximation where the lowest order 
expressions are compared with a set of low-energy data points, electric
charge, neutral currents and $\mu$-decay:
\begin{equation}
g^2 = 4\,\pi\alpha(0)\hsp, 
\qquad \mrM^2 = \frac{g^2\sqrt{2}}{8\mrG_\mrF} \hsp, 
\qquad 
\stws \quad{\mbox{from}} \quad
R = \frac{\sigma_{{\overline{\nu}} e}}{\sigma_{\nu e}} \hsp.
\end{equation}
The values for $g^2,\mrM^2$ and $\stws$, correct in the lowest order, will 
subsequently be used in the expressions for the radiative corrections.
In the next order we replace
\begin{equation}
g^2 \to g^2\lpar 1 + \delta g^2\rpar \hsp, \qquad
\mrM^2 \to \mrM^2\lpar 1 + \delta\mrM^2\rpar \hsp, \qquad
\stws\to \stws\lpar 1 + \delta \stws\rpar \hsp.
\end{equation}
The counter-terms are chosen to compensate precisely for the radiative
corrections for $ e$--$\mu$ scattering, $\mu$-decay and the ratio $R$.
Having determined these quantities we may proceed to making predictions.
The ratio $R$ is $R\lpar \stw \rpar$. We use the fact that it does not
depend on $g^2$ or $\mrM^2$ at lowest order and at zero energy and
momentum transfer. Thus
\begin{equation}
R_0\lpar \stws+ \stws\delta \stws\rpar = R_0\lpar \stws\rpar
+ R'_0\, \stws\delta \stws\hsp,
\end{equation}
with $R' = dR/d \stws$. The one-loop radiative corrections to $R$ will be some
$R_1$, then $\delta \stws$ is fixed by
\begin{equation}
\delta \stws = - \frac{R_1}{R'_0 \stws} \hsp.
\end{equation}
The rest is standard and gives $ e^4\to e^4(\,1 + \delta e^4)$ and
\begin{equation}
\delta g^2 = \frac{1}{2}\delta e^4 - \delta \stws\hsp, 
\qquad
\delta \mrM^2 = \frac{1}{2}\lpar \delta_{\mu} - \delta^{\rm em}_{\mu}\rpar 
+ \delta g^2 \hsp.
\end{equation}
Let us denote by $ \mrs^2_{\nu e}$ the low-energy weak mixing angle defined through some
$R_{{\rm exp}}$. Then we can derive masses for the vector bosons in the
low-energy convention. They are given by
\begin{equation}
\mrM^2_{\sPW} = \frac{\pi\alpha}{\sqrt{2}\mrG_\mrF \mrs^2_{\nu e}}\hsp, 
\qquad
\mrM^2_{\sPZ} = \frac{\pi\alpha}{\sqrt{2}\mrG_\mrF \mrs^2_{\nu e} c^2_{\nu e}}\hsp.
\end{equation}
Let us consider the amplitude for $\barnm e^{-}\to\barnm e^{-}$.
In real life many different contributions should be considered,
but to illustrate some of the relevant points in the procedure
it is enough to limit the calculation to contributions
from the heavy quark doublet, ($ t$--$ b$), 
to the $\mrZ \mrZ$ and $\mrZ \gamma$ transitions. 

In this case, we obtain
\begin{equation}
A\lpar\barnm e^{-}\to\barnm  e^{-}\rpar = 
\lpar \frac{ig}{4\,\ctw}\rpar^2\gamma^{\mu}\,\gamma_{+}
\otimes \gamma_{\mu}\,\bigl( a + b\,\gamma^5\bigr) \hsp,
\end{equation}
where $\gamma^{\alpha}\otimes\gamma^{\alpha} = {\overline\nu}_{\mu}
\gamma^{\alpha}\barnm {\overline e}\gamma^{\alpha} e$, etc.
and
\begin{equation}
a = \Bigl[ 4\, \stws - 1 - \frac{g^2 \stws}{4\,\pi^2}\,
\frac{\Sigma_{\sPZ \gamma}( p^2)}{ p^2} \Bigr]\,
\Delta_{\sPZ}( p^2)\hsp,  
\qquad
b = -\Delta_{\sPZ}( p^2)\hsp,  
\end{equation}
with a propagator $\Delta_{\sPZ}$ given by
\begin{equation}
\Delta^{-1}_{\sPZ}( p^2) =  p^2 + \mrM^2_{_0} - \frac{g^2}{16\,\pi^2 \ctws}\,
\Sigma_{\sPZ\sPZ}( p^2) \hsp,
\end{equation}
where $\Sigma_{\sPZ\sPZ}$ is the $\mrZ\,$-boson self{-}energy and $\Sigma_{\sPZ \gamma}$ is the
corresponding transition. Then the total cross-section $\sigma_{\nu e}$ can be computed 
and the data point $R = \sigma_{{\overline\nu} e}/\sigma_{\nu e}$ used:
\begin{equation}
R = \frac{\xi_{\nu e}^2 - \xi_{\nu e} + 1}{\xi_{\nu e}^2 + \xi_{\nu e} + 1} \hsp, 
\qquad
\xi_{\nu e} = \frac{a}{b} \hsp,
\end{equation}
where we assume the approximation of zero momentum transfer.
Subtracting the terms involving UV poles
and introducing a mass scale $\mu$ we define the counter-term for 
$\stw$ and fix $\supMSB{\stw}$ to first order in $\alpha$. 
The whole renormalization procedure amounts to throwing away
infinities. If we subtract the terms involving UV poles then the $\overline{MS}$
redefinition of the parameters is obtained (including $\stw$), but
we could as well assign any finite value to $\tDdr$ (to be defined in \eqn{UVdef}) and check for
the independence of the physical quantities of $\tDdr$.
From the $ \PQt$--$ \PQb$ quark doublet we obtain
\begin{eqnarray}
\supMSB{ \stws} &=&  \mrs^2_{\nu e} + \frac{\alpha}{12\,\pi}\,\Bigl[
2\,\lpar 1 - \frac{8}{3}\, \mrs^2_{\nu e}\rpar \,
\ln\frac{ m^2_t}{\mu^2} + \lpar 1 - \frac{4}{3}\, \mrs^2_{\nu e}\rpar \,
\ln\frac{ m^2_b}{\mu^2} \Bigr] \hsp,  \nonumber\\
 \mrs^2_{\nu e} &=& \frac{1 - \xi_{\nu e}}{4}\hsp. 
\end{eqnarray}
As expected, there are no terms quadratic in the quark masses.

At this point we are ready to make predictions. Starting from $ p^2 = 0$
we can introduce an effective $ p^2$-dependent weak mixing angle, etc. 

There is an important lesson to learn: there are Lagrangian parameters (\myeg $\stw$), input
parameters (\myeg $\xi_{\nu e}$) and predictions (or ``pseudo{-}observables'' ). In the SM there
is no one{-}to{-}one correspondence between Lagrangian parameters and input parameters. 
It was different in the older days of QED where only two Lagrangian parameters are present, $e$ and $\mrm_{\mre}$.
In this case one uses $\mrm_{\exp}\,(1 + \Delta \mrm_{e})$ reflecting some vage intuition about the physical meaning of
the bare mass. The strategy to prescribe precisely what a Lagrangian parameter is offers a problem when there is no
unique experimental quantity that can play the role of defining the parameter.

At LEP $\mrs^2_{\nu e}$ will be replaced by a $\sin\theta^\mrf_{\eff}$, related to the vector and axial couplings of the
$\mrZ\,$-boson.
\section{LEP: the $\rho\,$-parameter and pseudo-observables}
One way to explain renormalization is to say that infinities are
unobservable and can thus be absorbed into the parameters of the Lagrangian.
What about potentially large effects in the renormalized theory? When are
they observable? For instance the $m^2_{\PQt}\,$-terms at
one-loop, which are there and show up in physical observables. 
Conversely, the $\mrM_{\sPH}$ dependence of one-loop radiative corrections was
another seminal contribution of Veltman and it was 
described by the screening theorem~\cite{Veltman:1977kh,Veltman:1994vm}:
the one-loop $\mrM_{\sPH}$-dependence in physical observables is only logarithmic. 
Terms proportional to $\mrM^2_{\sPH}$ are unobservable 
at the one-loop level (they start at two loops) and can be absorbed into the 
parameters of the SM Lagrangian, apart from the case of one-loop diagrams 
with external (on-shell) Higgs{-}lines. The $\rho\,$-parameter~\cite{Ross:1975fq,Einhorn:1981cy,vanderBij:1983bw} was born.

A comment is needed for the ``original'' $\rho\,$-parameter, defined as the ratio of the $\PW$ and $\PZ$ 
masses squared divided by $\ctws$. We start by introducing a bare quantity:
$\rho_0= \mws/(\ctws\mzs)$. The quantities appearing in this relation must be related to
experimental data. There is no ambiguity in what is meant by an experimental mass. There remains 
the experimental $\ctws$ which should be extracted from data (at the time, low{-}energy data).
In the original formulation, where everything was extracted from low-energy data $(p^2 = 0)$, a $\rho$ 
was introduced as
\begin{equation}
\rho = 1 + \frac{g^2}{16\,\pi^2\mws}\,
\Bigl[
\Sigma_{\sPW\sPW}(0) - \Sigma_{\sPZ\sPZ}(0)
\Bigr]
= 1 + \frac{g^2}{16\,\pi^2}\,\Delta\rho\hsp,
\label{rhovelt}
\end{equation}
where the $1$ in the r.h.s. of the equation is actually $\rho_{\rm bare}$ in
the SM. The $\rho$-parameter of \eqn{rhovelt} is finite  
and numerically very close to the experimental one,
in any scheme where the counterterms are prescribed. The main correction is due to the top quark mass.
Usually we do not attach any particular relevance to bare parameters,
only the renormalized Lagrangian predicts meaningful --- measurable --- quantities. 
However, the $\rho$-parameter of the SM plays a special
role. It is finite because of a residual symmetry which is nothing but
the usual isospin invariance.
Individual components in \eqn{rhovelt} are by themselves infinite, but the 
combination occurring in this equation for the $\rho$-parameter is finite, 
as it should be. 

The Nobel Prize in Physics $1999$ was awarded ``for elucidating the quantum structure of electroweak interactions 
in physics''' which was crucial for LEP.
At LEP we had the SM with one missing ingredient, therefore the strategy was:
test the SM hypothesis versus $\mrM_{\sPH}$,
introduce Pseudo{-}Observables (generalizations of the $\rho$ parameter~\cite{Veltman:1977kh}), fit them and derive 
limits on the Higgs boson mass~\cite{Bardin:1999gt}. 

Ideally, the strategy should have been to combine the results of the LEP experiments
at the level of the measured cross-sections and asymmetries - a goal
that has never been achieved because of the intrinsic complexity, given 
the large number of measurements with different cuts and the complicated 
structure of the experimental covariance matrices relating their errors. This reflects computing limitations at the time.

What the experimenters did~\cite{Bardin:1999gt} was just collapsing (and/or transforming)
some ``primordial quantities'' (say number of observed events in some
pre-defined set-up) into some ``secondary quantities'' (the POs) which are
closer to the theoretical description of the phenomena.
The practical attitude of the experiments was to stay with a 
fit from ``primordial quantities'' to POs (with a SM remnant) for
each experiment, and these sets of POs were averaged.  
The result of this procedure are best values for POs. The extraction of Lagrangian
parameters, was based on the LEP-averaged POs. The PO{-}strategy was made possible thanks to high{-}precision
QED calculations and tools, from the pioneering work of \Bref{Berends:1987ab}(for an update see
\Bref{Blumlein:2020jrf}) up to the Monte Carlos 
that were instrumental at LEP in extracting the realistic observables; for instance 
\textit{KKMC}, the most advanced MC for $2\,$-fermion production~\cite{Jadach:1999vf} (including second order QED 
with resummation, initial{-}final state interference and spin polarizations, for an update see \Bref{Arbuzov:2020coe}).
\section{LHC: after the discovery of the Higgs boson}
After the LHC Run 1, the SM has been completed, raising its status to that of a full theory~\cite{Mariotti:2016owy}. 
Despite its successes, 
this SM has shortcomings vis-\`a-vis cosmological observations.
At the same time, there is presently a lack of direct evidence for new physics phenomena at the accelerator energy
frontier. From this state of affairs arises the need for a consistent theoretical framework in which deviations from the
SM predictions can be calculated. 

Theoretical physics suffers from some inherent difficulties: 
great successes during the $2$0th century, increasing difficulties to do better, as the easier 
problems get solved. The lesson of experiments from $1973$ to today is that it is extremely difficult to find a 
flaw in the SM~\footnote{Note however the long-standing tension between experiment and SM prediction in the 
anomalous magnetic moment of the muon, recently reaffirmed by the Fermilab experiment,
T.~Albahri et al. Muon $g - 2$, Phys. Rev. Accel. Beams (2021).}: maybe the SM includes 
elements of a truly fundamental theory. 
The conventional vision is: some very different physics occurs at Planck scale, the SM is just 
an effective field theory (EFT). What about the next SM? A new weakly{-}coupled renormalizable model? 
A tower of EFTs? Of course there is a different vision: is the SM close to a fundamental theory?

We need a consistent theoretical framework in which deviations from the SM (or next{-}SM) predictions 
can be calculated. Such a framework should be applicable to comprehensively describe measurements in all sectors 
of particle physics: LHC Higgs measurements, past electroweak precision data (EWPD), etc.
Here we outline the strategy:

\begin{itemize}

\item Consider the SM augmented with the inclusion of higher dimensional operators 
(say theory $\mrT_1$); not strictly renormalizable. Although workable to all orders, $\mrT_1$ 
fails above a certain scale, $\Lambda_1$.

\item Consider any beyond-standard-model (BSM) model that is strictly renormalizable and respects unitarity 
($\mrT_2$); its parameters can be fixed by comparison with data, while masses of heavy 
states are presently unknown. $\mrT_1 \not= \mrT_2$ in the UV but they must have the same IR
behavior.

\item Consider now the whole set of data below $\Lambda_1$. 
$\mrT_1$ should be able to explain them by fitting Wilson coefficients, 
$\mrT_2$ adjusting the masses of heavy states (as SM did with the
         Higgs mass at LEP) should be able to explain the data. 
Goodness of both explanations is crucial in understanding how well they match and how 
reasonable is to use $\mrT_1$ instead of the full $\mrT_2$.

\item Does $\mrT_2$ explain everything? Certainly not, but it should be able to 
explain something more than $\mrT_1$. 

\item We could now define $\mrT_3$ as $\mrT_2$ augmented with (its own) 
higher dimensional operators; it is valid up to a scale $\Lambda_2$.

\item Continue.

\end{itemize}

This prompts the important question whether there is a last fundamental theory in this tower 
of EFTs that supersede each other with rising energies~\footnote{Kuhlmann, Meinard, ``Quantum Field Theory'', 
The Stanford Encyclopedia of Philosophy (Winter 2018Edition), Edward N. Zalta (ed.)}. Some people conjecture that this 
deeper theory could be a string theory, \myie a theory which is not a field theory any more. 
Or should one ultimately expect from physics theories that they are only valid as approximations and in a 
limited domain? To summarize: experiments occur at finite energy and measure 
$\mrS^{\mathrm{eff}}(\Lambda)$ (an effective $\mrS\,$-matrix);
whatever QFT should give low energy
$\mrS^{\mathrm{eff}}(\Lambda)\,,\;\forall\,\Lambda < \infty$, \myie
there is no fundamental scale above which 
$\mrS^{\mathrm{eff}}(\Lambda)$ is not defined.
However $\mrS^{\mathrm{eff}}(\Lambda)$ loses its predictive power
if a process at $E = \Lambda$ requires 
an infinite number of renormalized parameters~\cite{Preskill:1990fr,Costello2011,Hartmann:2001zz}. 

To summarize: before LHC we had the SM, a weakly coupled, strictly renormalizable (a theory with $n$
Lagrangian parameters, requiring $n$ data points, requiring $n$ calculations; the
$(n+1)\,$th calculation is a prediction) theory with one unknown, $\mrM_{\sPH}$. 
The strategy was: test data against predictions vs. $\mrM_{\sPH}$.
At LHC, after the discovery, with  a lack of direct evidence for new physics phenomena
we have all the ingredients required to asses (in)consistency of the SM against data.

We briefly review the SMEFT Lagrangian~\cite{Ghezzi:2015vva,Passarino:2016pzb,Brivio:2017vri}: consider the 
standard model, described by a Lagrangian $\Lag^{(4)}_{\mySM}$ with 
a symmetry group $\mrG = SU(3)\,\times\,SU(2)\,\times\,U(1)$. The SMEFT extension is described by a Lagrangian
\begin{equation}
\Lag_{\SMEFT} = \Lag^{(4)}_{\mySM} + \sum_{d > 4}\,\sum_i\,\frac{\mra^d_i}{\Lambda^{4-d}}\,\Ope^{(d)}_i \hsp,
\label{SMEFTL}
\end{equation}
where $\Lambda$ is the cutoff of the effective theory, $\mra^d_i$ are Wilson coefficients and $\Ope^{(d)}_i$ are
$\mrG\,$-invariant operators of mass-dimension $d$ involving the $\Lag^{(4)}_{\mySM}$ fields.
In this work we will use the so{-}called ``Warsaw basis''~\cite{Grzadkowski:2010es}.

Unconventional approach to EFT: derivative{-}coupled field theories are known to develop ghosts~\cite{Ostro}. 
The EFT option~\cite{Passarino:2019yjx} replaces
the original $\Lag$ with some $\Lag_{\eff}$ truncated at some order in the $\Lambda\,$-expansion; the ``dangerous''
terms are substituted by using the equations of motion where, for instance, we neglect terms of $\mcO(\Lambda^{-2})$. We
assume that $\Lag_{\eff}$ will be replaced by a well{-}behaved $\Lag^{\prime}$ at some larger scale, therefore justifying 
a truncated perturbative expansion; the EFT does not have ghosts while remaining within its regime of validity. 

\vspace{0.2cm}
\noindent
{\itshape{SMEFT and renormalization}} \hspace{0pt} \\
At this point we have lost strict renormalizability but this should not come at the price of loosing
computability~\cite{David:2015waa,David:2020pzt}; 
whether the predictions of a theory are matched by Nature is a completely different matter and can be decided only 
by comparing the predictions with experiment (calculability).

A renormalizable theory is determined by a fixed number 
of parameters; once these are determined (after finite renormalization) we can make definite predictions at a 
fixed accuracy. An EFT theory requires at higher and higher energies more and more counterterms; the asymptotic 
expansion in $\mrE/\Lambda$ may break down completely above some scale. Given a truncated expansion, we still have a 
large family of UV{-}complete theories with these low order terms, which have different behavior at higher energies. 
Note that the notion of UV completion adopted here~\cite{Crowther:2017pho} is the claim that a theory is ``formally'' 
predictive up to all (possible) high energies, but we do not include the additional criterion that the theory be 
a final, unified ``theory of everything''.

Therefore, it is crucial to prove that our EFT is closed under renormalization, order-by-order in the asymptotic 
expansion, although the number of counterterms will grow with the order (as mentioned above, the predictive power 
is lost at scales approaching the cutoff). For any given process the amplitude can be written as follows:
\begin{eqnarray}
\mrA &=& \sum_{n=\mrN}^{\infty}\,g^n\,\mrA^{(4)}_n +
       \sum_{n=\mrN_6}^{\infty}\,\sum_{l=1}^n\,\sum_{k=1}^{\infty}\,
        g^n\,g^l_{4+2\,k}\,
        \mrA^{(4+2\,k)}_{n\,l\,k} \hsp,
\end{eqnarray}
where $g$ is the $SU(2)$ coupling constant and 
$g_{4+2\,k} = 1/(\sqrt{2}\,\mrG_{\mrF}\,\Lambda^2)^k = g^k_6$,
where $\mrG_{\mrF}$ is the Fermi coupling constant and $\Lambda$ is the scale around which
new physics (NP) must be resolved.
For each process, $N$ defines the $\mrdim = 4$ LO (\myeg $N = 1$ for $\mrH \to \mrV\mrV$
etc. but $N = 3$ for $\mrH \to \gamma\gamma$). $N_6 = N$ for tree initiated processes and
$N - 2$ for loop initiated ones. Here we consider single insertions of $\mrdim = 6$ operators,
which defines the so{-}called NLO SMEFT.
To be more precise, we define a NLO SMEFT amplitude as the one containing
SMEFT vertices inserted in tree-level SM diagrams,
tree-level (SMEFT-induced) diagrams with a non-SM topology,
SMEFT vertices inserted in one-loop SM diagrams, and
one-loop (SMEFT-induced) non-SM diagrams. 

The amplitude can be rewritten as
\begin{eqnarray}
\mrA &=&
g^N\,\mrA^{(4)}_{\myLO}\bigl(\{p\}\bigr) +
g^N\,g_{_6}\,\mrA^{(6)}_{\myLO}\bigl(\{p\}\,,\,\{\mra\}\bigr) +
\frac{g^{N+2}}{16\,\pi^2}\,\mrA^{(4)}_{\myNLO}\bigl(\{p\}\bigr) 
\nonumber \\
{}&+&
\frac{g^{N+2}\,g_6}{16\,\pi^2}\,\mrA^{(6)}_{\myNLO}\bigl(\{p\}\,,\,\{\mra\}\bigr) \hsp,
\end{eqnarray}
where $\{p\}$ is the set of SM parameters and $\{\mra\}$ the set of Wilson coefficients.
Counterterms are introduced using
\begin{equation}
\DUV = \frac{2}{4 - \mrd} - \gamma - \ln \pi - \ln\frac{\mu^2_{\mrR}}{\mu^2} = 
\frac{1}{\bar{\varepsilon}} - \ln\frac{\mu^2_{\mrR}}{\mu^2} \hsp,
\label{UVdef}
\end{equation}
where $\mrd$ is the space-time dimension, $\gamma$ is the Euler{-}Mascheroni constant, 
the loop measure is $\mu^{4 - d}\,d^dq$ and $\mu_{\mrR}$ is the renormalization scale.
The counterterms are defined by
\begin{equation}
\mrZ_i = 1 + \frac{g^2}{16\,\pi^2}\,\bigl( 
d\mrZ^{(4)}_i + g_6\,d\mrZ^{(6)}_i \bigr)\,\DUV \hsp.
\end{equation} 
With field/parameter counterterms we can make UV finite (at $\mcO(g^2\,g_6)$) all self{-}energies and transitions 
and the corresponding Dyson resummed propagators. Of course we have to prove cancellation of UV poles for all
Green's functions. This means that we need to make the SMEFT $\mrS\,$-matrix UV (and IR)
finite, including $\mrdim = 6$ operators and, at least, $\mrdim = 8$ operators (truncation uncertainty).
The verification of any claim with explicit computations is of importance. 
The role of symmetry is crucial. The best way to understand the connection between UV poles and symmetry is given by
the background{-}field{-}method~\cite{tHooft:1973bhk,tHooft:1974toh}. Let us give an example of the complexity of 
proving cancellation of UV poles in any EFT.
Consider a scalar theory
\begin{eqnarray}
\Lag(\phi_{\mrc} + \phi) &=&
\Lag(\phi_{\mrc}) + \phi_i\,\Lag'_i(\phi_{\mrc}) + 
\frac{1}{2}\,\partial_{\mu}\,\phi_i\,\partial_{\mu}\,\phi_i 
\nonumber\\
{}&+& \phi_i\,\mrN^{\mu}_{ij}\,\partial_{\mu}\,\phi_j +
\frac{1}{2}\,\phi_i\,\mrM_{ij}(\phi_{\mrc})\,\phi_j + \mcO(\phi^3) +
\mbox{tot. der.}
\end{eqnarray}
($\Lag'_i(\phi_{\mrc}) = 0$).
All one loop diagrams are generated by $\Lag_2(\phi)$, the part quadratic in $\phi$.
\begin{eqnarray}
\Lag_2(\phi) &\to& - \frac{1}{2}\,\lpar \partial_{\mu}\,\phi \rpar^2 + 
\phi\,\mrN^{\mu}\,\partial_{\mu}\,\phi +
\frac{1}{2}\,\phi\,\mrM\,\phi \hsp.
\end{eqnarray}
The counter{-}Lagrangian is given by
\begin{eqnarray}
\Delta\,\Lag  &=& \frac{1}{8\,\pi^2\,(d-4)}\,\Bigl[
a_0\,M^2 + a_1\,\lpar \partial_{\mu}\,\mrN_{\nu} \rpar^2 +
a_2\,\lpar \partial_{\mu}\,\mrN_{\mu} \rpar^2 +
a_3\,\mrM\,\mrN^2 
\nonumber\\
{}&+&
a_4\,\mrN_{\mu}\,\mrN_{\nu}\,\partial_{\mu}\,\mrN_{\nu} +
a_5\,\lpar \mrN^2 \rpar^2 +
a_6\,\lpar \mrN_{\mu}\,\mrN_{\nu}\rpar^2
\Bigr] \hsp.
\end{eqnarray}
However, define $X = \mrM - \mrN^{\mu}\,\mrN_{\mu}$ and see that $\Lag$ is invariant under
't Hooft transformation ($\mrH$) , $\Lambda$ antisymmetric 
\begin{eqnarray}
%
& \phi' = \phi + \Lambda\,\phi \hsp,
\quad
\mrN'_{\mu} = \mrN_{\mu} - \partial_{\mu}\,\Lambda + \bigl[ \Lambda\,,\,\mrN_{\mu} \bigr] \hsp,
\quad
X^{\prime} = X + \bigl[ \Lambda\,,\,X \bigr] &
\end{eqnarray}
Therefore $\Delta\,\Lag$ also will be invariant ($\mathrm{Tr}\,X$ is invariant)
\begin{eqnarray}
\Delta\,\Lag &=& \frac{1}{\varepsilon}\,\mbox{Tr}\,
\lpar a\,X^2 + b\,Y^{\mu\nu}\,Y_{\mu\nu} \rpar \hsp, 
\nonumber\\
Y_{\mu\nu} &=& \partial_{\mu}\,\mrN_{\nu} -
\partial_{\nu}\,\mrN_{\mu} +
\bigl[ \mrN_{\mu}\,,\,\mrN_{\nu} \bigr] \hsp.
\label{CTL}
\end{eqnarray}
and $Y$ transforms as $X$. The counter{-}Lagrangian is made of products of objects transforming as $X$ and of
$\mrdim = 4$.
As a consequence of the $\mrH\,$-invariance the number of counterterms goes from $7$ to $2$.
Any approach to SMEFT violating invariance is doomed to failure~\cite{Passarino:2016saj}. 
An EFT (\myeg SMEFT) including $\mrdim = 6,8$ operators will contain a term (for $\mrdim = 6$ see \Bref{Buchalla:2019wsc})
\begin{eqnarray}
& \frac{1}{2}\,\partial_{\mu}\,\phi_i\;\;
\mrg^{\mu\nu}_{ij}(\phi_{\mrc}) 
\;\;\partial_{\nu}\,\phi_j &
\nonumber\\
{}& \mbox{matrix{-}valued metric tensor} &
\label{metric}
\end{eqnarray} 
We should pay attention to the fact that in the SM $\mrg^{\mu\nu}_{ij} \propto \delta^{\mu\nu}\,\delta_{ij}$ while in 
quantum gravity (QGR) it remains diagonal only in the $ij$ indices. 
Therefore, in SMEFT we will have matrix{-}valued Riemann tensors, \myie more invariants for the counter{-}Lagrangian 
$\Delta\,\Lag$, \myie 
$\mathrm{Tr}(X\,\mrR), \mathrm{Tr}\,(\mrR^2)\,\dots$; as a consequence, EFT is computationally more complex than 
QGR~\cite{Veltman:1975vx}. 
We should remember that the name 
of the game is to have the full $\Delta\,\Lag$, not the counterterms for one or two processes.
If $\Lag$ is invariant under a group $\mrG$ then the relation between the $\mrG$ transformation and 
the $\mrH$ one is crucial in proving closure under renormalization (not the same as strict renormalizability).
In other words the $\mrH\,$-invariant counterterms of \eqn{CTL} must also be $\mrG\,$-invariant.

To give an example of \eqn{metric} we consider a Lagrangian
\begin{equation}
\Lag= - \partial_{\mu} \Phi^*\,\partial_{\mu} \Phi  - \mrm^2\,\Phi^* \Phi - \frac{1}{2}\,g^2\,\bigl( \Phi^* \Phi \bigr)
\hsp,
\end{equation}
where $\Phi = \bigl( \phi_1 + \mrv + i\,\phi_2\bigr)/\sqrt{2}$. We introduce $v = \mrM/g$ and 
$\mrm^2 = \beta - \mrM^2/2$ and add
\begin{equation}
\Ope^{(8)} = 
\Bigl( \frac{g g_6}{\mrM^2} \Bigr)^2\,\partial_{\mu} \Phi^*\,\partial_{\mu} \Phi\,\Box\,\bigl( \Phi^* \Phi \bigr) \hsp.
\end{equation}
After splitting $\phi_i = \phi_{i\mrc} + \phi_i$ we obtain that the contribution of the operator of $\mrdim = 8$
to the metric tensor is $\delta g = \delta_1 g + \delta_2 g + \mcO(g^3)$, where
\begin{equation}
\delta_1 g^{\mu\nu} = \frac{g g^2_6}{\mrM}\,\phi_{1\mrc}\,\delta^{\mu\nu}\,\mrI \hsp,
\end{equation}
{\footnotesize{
\[
\delta_2 g^{\mu\nu} = \frac{g^2 g^2_6}{\mrM^4} \left(
\begin{array}{ll}
(g_{11} + g_{22})\,\delta^{\mu\nu} + 4\,\partial^{\mu} \phi_{1\mrc}\,\partial^{\nu} \phi_{1\mrc} &
4\,\partial^{\mu} \phi_{1\mrc} \partial^{\nu} \phi_{2\mrc} \\
4\,\partial^{\mu} \phi_{12\mrc} \partial^{\nu} \phi_{1\mrc} &
(g_{11} + g_{22})\,\delta^{\mu\nu} + 4\,\partial^{\mu} \phi_{2\mrc}\,\partial^{\nu} \phi_{2\mrc} 
\end{array}
\right)
\]
}}

\noindent
where $g_{ij}= \partial^{\alpha} \phi_{i\mrc}\,\partial^{\alpha} \phi_{j\mrc}$.

Here we will summarize the main steps in the renormalization procedure for the SMEFT. 
Field/parameter counterterms are not enough to make UV finite the Green's functions with more 
than two legs. A mixing matrix among Wilson coefficients is needed:
\begin{equation}
\mra_i = \sum_j\,\mrZ^{\sPW}_{ij}\,\mra^{\mathrm{ren}}_j \hsp, 
\quad 
\mrZ^{\sPW}_{ij} = \delta_{ij} + \frac{g^2}{16\,\pi^2}\,d\mrZ^{\sPW}_{ij}\,\DUV \hsp.
\end{equation}
We can start by computing the amplitude for the (on{-}shell) decay $\mrH \to \gamma \gamma$ and fix a few
entries in the mixing matrix; we can continue with the $\mrH \mrZ \gamma$ and $\mrH \mrZ \mrZ$ amplitudes.
When we arrive at $\mrH \mrW \mrW$ we find that the $\mrdim = 4$ part can be made UV finite but for the $\mrdim = 6$
part there are no Wilson coefficients left free so that the UV finiteness follows from gauge cancellations. 
We then continue with the decay of the Higgs boson (and of the $\mrZ$ boson) into fermion pairs and make all the
corresponding amplitudes UV finite.

At this point we are left with the universality of the electric charge. In QED there is a
Ward identity telling us that $e$ is renormalized in terms of vacuum polarization,
Ward-Slavnov-Taylor identities allow us to generalize the argument to the full SM~\cite{Actis:2006rc,Dittmaier:2021loa}.

We can give a quantitative meaning to the the previous statement by saying that
the contribution from vertices (at zero momentum transfer) exactly cancel those from
(fermion) wave function renormalization factors. Therefore,
we need to
compute the vertex ${\overline{f}} f \gamma$ (at $q^2 = 0$) and the $f$ wave function 
factor in SMEFT, proving that the WST identity can be extended 
to $\mrdim = 6$; this is non trivial since there are no free Wilson coefficients in these terms (after the previous steps);
(non-trivial) finiteness of LEP processes follows.

Finite renormalization in SMEFT; let us recall that the renormalization procedure comprises the specification of 
the gauge{-}fixing term including, together with the corresponding FP Lagrangian, the choice of the 
regularization scheme --- nowadays dimensional regularization --- the prescription for the renormalization scheme and 
the choice of a input parameter set.
For the SM parameters we use on{-}shell renormalization which requires the choice of
some input parameter set (IPS); for the Wilson coefficients we use the $\overline{MS}$ scheme. Therefore, the final 
answer will contain ``universal'' logarithms (renormalization group) but also ``non{-}universal'' logarithms which
depend on the choice of the IPS. The choice of the IPS should be such that the effect of ``non{-}universal'' logarithms
is minimal, \myeg $\alpha(0)$ should be avoided. Once we work with NLO SMEFT, non{-}local effects (\myeg
normal{-}threshold singularities or even anomalous thresholds~\cite{Passarino:2018wix}) will also show 
up~\cite{delAguila:2016zcb,Donoghue:2017pgk,Passarino:2019yjx}
for all those observables where the light masses are small compared
to the scale at which we test the process and much smaller than the cutoff $\Lambda$.

Finally, the infrared/collinear part of the one-loop virtual and of the real corrections shows double factorization
and the total is finite~\cite{Boggia:2017hyq} at $\mcO(g^4\,g_6)$.

\vspace{0.2cm}
\noindent
{\itshape{Asymptotics of BSM models}} \hspace{0pt} \\
The backround-field-method is also used in combination with the heat kernel~\cite{Vassilevich:2003xt}, a very convenient 
tool for studying 
various asymptotics of the effective action; for instance in deriving the low{-}energy limit of some underlying BSM theory
described by $\Lag(\{\Phi\})$, where $\{\Phi\}$ includes both heavy and light fields. We expand
$\Phi= \Phi_{\mrc} + \phi$ and derive
\begin{eqnarray}
\Lag &=& \Lag_c + \langle\,\phi\,,\,\mrD \phi\,\rangle = \Lag_c + \phi^{\dagger}\,\mrQ\,\phi\hsp,
\nonumber \\
\mrZ\bigl[ \Phi_{\mrc} \bigr] &=& \int \bigl[ \mathcal{D} \phi \bigr]\,   
\exp\{ i\,\mrS \} = \exp\{i\,\mrS_c\}\,\mathrm{det}^{-1/2}(\mrD) \hsp,
\end{eqnarray}
where $\mrD$ must be self{-}adjoint. Using $\mrQ= \Box - \mrM + {\hat{\mrQ}}(\Phi_{\mrc})$ 
the heat kernel expansion requires computing
\begin{equation}
\mathrm{Tr}\,\ln \mrQ(x)\,\delta^4(x - y) =
\int d^4x\,\frac{d^4q}{(2\,\pi)^4}\,\mathrm{tr}\,\ln \Bigl[
- q^2 - M + \Box + 2\,i\,\spro{q}{\partial} + {\hat{\mrQ}(x\,,\,\partial_x)} \Bigr] \hsp.
\end{equation}
When there is one field or $\mrM_{ij} = \mrM^2\,\delta_{ij}$ we write
$\ln \Bigl[ - (q^2 + \mrM^2)\,(\mathrm{I} + \mrK) \Bigr]$,
expand $\ln (\mathrm{I} + \mrK)$ in powers of $\mrK$ obtaining the large $\mrM${-}expansion of 
$\mrS_{\eff} = 1/2\,\ln\,\mathrm{det}(\mrD)$ in terms of tadpole integrals.
Otherwise, with more heavy scales or mixed heavy{-}light scales,   
the correct Taylor expansion~\cite{Lashkari:2018tjh} is 
\begin{equation}
\ln (\mrA + \mrB) - \ln \mrA = \int_0^{\infty}\,d\mu^2\,\Bigl[
\mrA^{-1}_+\,\mrB\,\mrA^{-1}_+ -
\mrA^{-1}_+\,\mrB\,\mrA^{-1}_+\,\mrB\,\mrA^{-1}_+ \;+\, \dots \Bigr] \hsp, 
\end{equation}
where $\mrA_+ = \mrA + \mu^2\,\mathrm{I}$.
\section{Advanced SMEFT}
In this Section we will use the so{-}called ``Warsaw basis''~\cite{Grzadkowski:2010es}; note however that we will rescale 
the Wilson coefficients: in front of an operator $\Ope^{(\mrd)}_i$ of dimension $\mrd$ containing $n$ fields we will write
$g^{n-2}\,\mra^{\mrd}_i/\Lambda^{\mrd-4}$, where $g$ is the $SU(2)$ coupling constant.
Our goal is to consider the SMEFT with its anomalies~\cite{Preskill:1990fr,Cata:2020crs,Bonnefoy:2020tyv,Feruglio:2020kfq}; 
in these calculations a certain amount of $\gamma\,$-matrix
manipulation is unavoidable and we must specify the regularization scheme to be used, in particular, we must specify 
how to treat $\gamma^5$. We will use the scheme developed by Veltman in \Bref{Veltman:1988au}, which is based
on the work of \Brefs{tHooft:1972tcz,Breitenlohner:1975hg}. 
Therefore, $\gamma^{\mu}, \gamma^5$, and $\varepsilon^{\mu\nu\alpha\beta}$ are formal objects where
\begin{eqnarray}
\bigl\{ \gamma^{\mu}\,,\,\gamma^{\nu} \bigr\} = 2\,\delta^{\mu\nu}\,\mrI  & \quad & \mathrm{Tr}\,\mrI = 4 \hsp,
\nonumber \\  
\delta^{\mu\nu} = \delta^{\bar{\mu}\bar{\nu}} + \delta^{\hat{\mu}\hat{\nu}}, \;\;
\delta^{\bar{\mu}\bar{\mu}} = 4, & \quad &
\delta^{\hat{\mu}\hat{\mu}} = \mrd - 4, 
\nonumber \\
\delta^{\mu\alpha}\,\delta^{\alpha\nu} &=& 
\delta^{\bar{\mu}\bar{\alpha}}\,\delta^{\bar{\alpha}\bar{\nu}} + 
\delta^{\hat{\mu}\hat{\alpha}}\,\delta^{\hat{\alpha}\hat{\nu}} \hsp, 
\end{eqnarray}
where $\mrd$ is the space-time dimension. As described in \Bref{Veltman:1988au} we have the following relation
\begin{eqnarray}
\mathrm{Tr}\,\bigl( \cdots \,\gamma^{\alpha} \,\cdots \,\gamma^{\beta} \,\cdots \bigr) &=&
\mathrm{Tr}\,\bigl( \cdots \,\gamma^{\bar\alpha} \,\cdots \,\gamma^{\bar\beta} \,\cdots \bigr) 
\nonumber \\
{}&+&
\mathrm{Tr}\,\bigl( \cdots \,\gamma^{\hat\alpha} \,\cdots \,\gamma^{\hat\beta} \,\cdots \bigr) \hsp,
\label{anomTr}
\end{eqnarray}
where the dots indicate strings of four{-}dimensional gamma matrices and also $\gamma^5$. The second trace in the
r.h.s. of \eqn{anomTr} is computed according to the following rules:
\begin{enumerate}

\item move all the $\gamma^{\hat\mu}$ matrices to the right using $\gamma^{\hat\mu}\,\gamma^{\bar\nu} = -
\gamma^{\bar\nu}\,\gamma^{\hat\mu}$,

\item for a trace containing an odd number of $\gamma^5$ matrices use $\gamma^{\hat\mu}\,\gamma^5 = \gamma^5\,\gamma^{\hat\mu}$,

\item for a trace containing an even number of $\gamma^5$ matrices use 
$\gamma^{\hat\mu}\,\gamma^5 = - \gamma^5\,\gamma^{\hat\mu}$.

\end{enumerate}
As a consequence we obtain
\begin{equation}
\mathrm{Tr}\,\bigl( \cdots \,\gamma^{\hat\alpha} \,\gamma^{\hat\beta} \,\cdots \bigr) =
\mathrm{Tr}\,\bigl( \cdots \bigr)\,\mathrm{Tr}\,\bigl(\gamma^{\hat\alpha} \,\gamma^{\hat\beta}\bigr) \hsp,
\end{equation}
where the first trace in the r.h.s. only contains four-dimensional quantities while
$\mathrm{Tr}\,\bigl(\gamma^{\hat\alpha} \,\gamma^{\hat\beta}\bigr) = \mrd - 4$ \etc

Using a $\gamma^5$ which does not anticommute with the other Dirac matrices leads to ``spurious anomalies'' which 
violate gauge invariance (spoiling renormalizability) and we must impose the relevant 
Ward-Takahashi (\Brefs{Ward:1950xp,Takahashi:1957xn})
and Slavnov-Taylor (\Brefs{Veltman:1970dwc,Slavnov:1972fg,Taylor:1971ff}) 
identities (for a review see \Bref{Jegerlehner:2000dz}). The problems of course are related to the existence of the 
ABJ anomaly (\Brefs{Adler:1969gk,Bell:1969ts,Bardeen:1969md}), which cancels in the SM. The goal of this Section is to study 
the ABJ anomaly in the SMEFT extension of the SM.
\subsection{SMEFT and WTST identities \label{ident}}
Consider the following amplitudes:
\begin{eqnarray}
{}&{}& \mrA^{\sPZ}_{\mu\,;\,\nu_1,\,\dots\,,\nu_n}(p\,;\,q_1\,\dots\,,q_n\,;\,k_1\,,\dots\,,k_m) \hsp,
\nonumber \\
{}&{}& \mrA^{\phi}_{\nu_1,\,\dots\,,\nu_n}(p\,;\,q_1\,\dots\,,q_n\,;\,k_1\,,\dots\,,k_m) \hsp,
\end{eqnarray}
involving a $\PZ\,$-boson (or a $\phi^0$ Higss{-}Kibble ghost~\cite{Kibble:1967sv}) of momentum $p$, $n$ 
gauge bosons ($\PA, \PZ, \PW$) and $m$ Higgs bosons (all momenta are flowing inwards). The corresponding WTST identity is
\begin{equation}
\Gamma_{\nu_1,\,\dots\,,\nu_n} \;=\; 
i\,p^{\mu}\,\mrA^{\sPZ}_{\mu\,;\,\nu_1,\,\dots\,,\nu_n} + \mrM_{\sPZ}\,\mrA^{\phi}_{\nu_1,\,\dots\,,\nu_n} \;=\; 0 \hsp.
\label{WTSTi}
\end{equation}
We will study $2$ different schemes:
\begin{itemize}

\item[S1)] $\PZ$ and $\phi^0$ off{-}shell, remaining gauge bosons coupled to physical sources, \myie
$\partial_{\mu}\,\mrJ_{\mu} = 0$, anti{-}commuting $\gamma^5$;

\item[S2)] $\PZ$ and $\phi^0$ off{-}shell, remaining gauge bosons coupled to physical sources, \myie
$\partial_{\mu}\,\mrJ_{\mu} = 0$, Veltman{-}prescription for $\gamma^5$, four-dimensional, on{-}shell, external momenta.

\end{itemize} 

It is worth noting that there is a third scheme (which will not be considered here): $\PZ$ and $\phi^0$ off{-}shell, 
remaining gauge bosons coupled to arbitrary sources, \myie
$\partial_{\mu}\,\mrJ_{\mu} \not= 0$, Veltman{-}prescription for $\gamma^5$, $\mrd$-dimensional, off{-}shell external 
momenta, \myie $p_{\mu} = p_{\bar{\mu}} + p_{\hat{\mu}}$.

Consider the SMEFT, a theory which is not strictly renormalizable; if there were no anomalies, all the UV divergences could 
be cancelled in $\mrS^{\eff}(\Lambda)$, order{-}by{-}order in $1/\Lambda$. SMEFT loses its predictive power if a process at
$\mrE = \Lambda$ requires an infinite number of renormalized parameters.
Due to the anomaly at the one{-}loop level, the symmetry will be broken and the mechanism of cancelling divergences is 
disturbed, \myie WTST identities break down. For example one relevant identity concerns the amplitude for a $\PZ$ to
decay into two photons: if this identity is violated we can still restore it by introducing a UV{-}finite
counterterm as it would be the case in the SM with only an electron and a neutrino. However, this new term is of
non{-}renormalizable type giving rise to infinities at higher orders; see \Bref{Sterling:1981za} for a complete
discussion. The ``complete'' SM is an anomaly{-}free theory and we want to investigate the SMEFT from the following point of
view: are there anomalies in the SMEFT? If they show up, could we find relations among the Wilson coefficients that
cancel the anomalies? 
In the following sections we will consider the relevant WTST identities for SMEFT amplitudes
\begin{equation}
\mrA= \mrA^{(4)} + g_{_6}\,\mrA^{(6)} \hsp, \qquad g_{_6}= \frac{1}{\sqrt{2}\,\mrG_{\mrF}\,\Lambda^2} \hsp,
\label{ampli}
\end{equation}
where $\mrG_{\mrF}$ is the Fermi coupling constant and the first term in \eqn{ampli} is the SM contribution. In computing the 
WTST identity of \eqn{WTSTi} (schemes $1$ and $2$) we assume that all the sources for the $n$ gauge bosons and the 
$m$ Higgs bosons are physical and on{-}shell, \myie $\spro{q_i}{\mrJ_i(q_i)} = 0$; this means that external sources 
cannot absorb/emit Fadeev{-}Popov ghosts. Internal lines in one{-}loop diagrams represent fermions.
Each one{-}loop amplitude can be decomposed as follows:
\begin{eqnarray}
\mrA &=& \frac{\mrS}{\mrd - 4} + \mrR + \sum_{\mra}\Pf_{\mra}\,\mrA^{\fin}_0(\mra) +
\sum_{\mrb}\Pf_{\mrb}\,\mrB^{\fin}_0(\mrb) 
\nonumber \\
{}&+&
\sum_{\mrc}\Pf_{\mrc}\,\mrC_0(\mrc) +
\sum_{\mrd}\Pf_{\mrd}\,\mrD_0(\mrd)  \hsp,
\end{eqnarray}
where $\mrA_0,\,\dots\,\mrD_0$ are scalar one,$\,\dots\,$four point functions~\cite{tHooft:1978jhc,Passarino:1978jh}, 
$\mrd$ is the space{-}time dimension and ``fin'' denotes the UV finite part. It is worth noting that anomalies can be 
cancelled by adding counterterms if and only if the anomaly is of $\mrR$ type, \myie UV finite and local. Locality of 
the UV{-}finite counterterms is related to the unitarity of the theory.
\subsection{Anomalies and anomalous terms} 
Anomalies and anomalous terms, although correlated, are not the same thing. Anomalies have to do with WTST identities,
perhaps the best example is given by
\begin{equation}
\Gamma_{\alpha\beta} = i p^{\lambda}\,\mrA^{\sPZ}_{\lambda\,;\,\alpha\,,\,\beta} +
\mrM_{\sPZ}\,\mrA^{\phi}_{\alpha\,,\,\beta} \hsp,
\end{equation}
the $\PZ \PA \PA$ WTST identity. If this identity is violated, say
\begin{equation}
\Gamma_{\alpha\beta} = \PX\,\varepsilon_{\mu\nu\alpha\beta}\,p^{\mu}_1\,p^{\nu}_2 \hsp,
\end{equation}
with $\PX$ UV{-}finite and local, we can restore it by adding to the Lagrangian a term 
$\PX\,\varepsilon_{\mu\nu\alpha\beta}\,\phi^0\,(\partial_{\mu}\,\PA_{\alpha})\,(\partial_{\nu}\,\PA_{\beta})$. One could think of
introducing two counterterms, $\PZ\PA\PA$ and $\phi^0\PA\PA$, in order to restore the identity and to cancel, at the same 
time, the anomalous $\PZ\PA\PA$ and $\phi^0\PA\PA$ couplings. However, these counterterms are not local. Having 
loop{-}induced anomalous couplings is not a surprise, even in the SM. The important thing is the cancellation of
anomalies, not the presence of anomalous couplings. The point was explained long ago in \Bref{Beenakker:1996kn};
consider any $\mrV \mrV \mrV$ one{-}loop, SM, vertex, it will contain $\varepsilon${-}terms that contribute to the triangle 
anomaly. The cancellation of the anomaly requires the contribution of the massive top quark. Writing
\begin{equation}
\mrV = \sum_{\Pf}\,\mrV_{\Pf}(m_{\Pf} = 0) + \Bigl[ \mrV_{\PQt}(m_{\PQt} \not= 0) - \mrV_{\PQt}(m_{\PQt} = 0) \Bigr] \hsp,
\end{equation}
it follows from the anomaly-cancellation conditions that all
$\varepsilon${-}terms disappear in the (massless) sum on the right-hand side. The remainder contains
$\varepsilon${-}terms and is $m_{\PQt}${-}dependent. This dependence is known to produce effects of delayed
unitarity cancellation, which may become relevant at high energies.
Everything becomes clear when we consider the explicit expressions in the SM for 
$\PZ_{\mu}(P) \to \PA_{\alpha}(p_1) + \PA_{\beta}(p_2)$ and
$\phi^0(P) \to \PA_{\alpha}(p_1) + \PA_{\beta}(p_2)$:
\begin{eqnarray}
\mrA^{\phi}_{\alpha\beta} &=& 
\frac{g^3}{12\,\pi^2}\,\frac{\stws}{\ctw}\,\varepsilon_{\mu\nu\alpha\beta}\,p^{\mu}_1\,p^{\nu}_2
\Bigl[
3\,\sum_{\Pl}\frac{m^2_{\Pl}}{M^2_{\sPZ}}\,\mrC_0(m_{\Pl}) 
\nonumber \\
{}&+&
\sum_{\PQd}\frac{m^2_{\PQd}}{M^2_{\sPZ}}\,\mrC_0(m_{\PQd}) -
4\,\sum_{\PQu}\frac{m^2_{\PQu}}{M^2_{\sPZ}}\,\mrC_0(m_{\PQu}) \Bigr] \hsp,
\nonumber \\ \\
\mrA^{\sPZ}_{\mu\,;\,\alpha\beta} &=&
- i\,\frac{g^3}{12\,\pi^2}\,\frac{\stws}{\ctw}\,\varepsilon_{\mu\nu\alpha\beta}\,p^{\mu}_1\,p^{\nu}_2
\Bigl[
3\,\sum_{\Pl}\frac{m^2_{\Pl}}{s}\,\mrC_0(m_{\Pl}) 
\nonumber \\
{}&+&
\sum_{\PQd}\frac{m^2_{\PQd}}{s}\,\mrC_0(m_{\PQd}) -
4\,\sum_{\PQu}\frac{m^2_{\PQu}}{s}\,\mrC_0(m_{\PQu}) \Bigr]\,P_{\mu} \hsp,
\end{eqnarray}
where $P = - p_1 - p_2$, $P^2= - s$ and $\mrC_0(m_{\Pf})$ is the scalar three{-}point function with internal $\Pf$ lines.
The sum is over leptons, up and down quarks.
The WTST identity is obviously satisfied but both amplitudes correspond to anomalous, non{-}local, couplings. As far as the
$\phi^0 \PA \PA$ amplitude is concerned this fact is not so relevant since $\phi^0$ is not an asymptotics state.
Furthermore, the $\PZ \PA \PA$ amplitude gives zero when the $\PZ$ boson is coupled to a conserved current.
To summarize, the no{-}anomaly scheme does not imply the absence of anomalous couplings. Cancellation of anomalous 
couplings is not the question, cancellation of anomalies is.
As we will discuss in the next paragraph, this WTST identity is violated in the SMEFT.
\subsection{Step $0$:  WTST identities for two{-}point functions}
For two{-}point functions scheme $1$ and scheme $2$ give the same result, \myie all WTST identities are satisfied, both
in the SM and in the SMEFT~\cite{Ghezzi:2015vva,Corbett:2020ymv}. It is worth noting that the identities are violated 
if we use a scheme where
$\{ \gamma^{\bar{\mu}}\,,\,\gamma^5 \} = 0$ and $[ \gamma^{\hat{\mu}}\,,\,\gamma^5 ] = 0$, \myie without using the anomalous trace
introduced in \Bref{Veltman:1988au}. Of course, they can always be restored by introducing local and UV{-}finite
counterterms even in the presence of evanescent terms~\cite{Bonneau:1980ya}. These terms are formally zero in the limit
$\mrd \to 4$ but their effect must be carefully analyzed due to the presence of UV poles.
\subsection{Step $1$:  WTST identities for three{-}point functions}
In our notation $g$ is the $SU(2)$ coupling constant, $\ctw = \mrM_{\sPW}/\mrM_{\sPZ}$ is the cosine of 
the weak{-}mixing angle. Furthermore, we introduce the following combinations,
\begin{eqnarray}
\alW = \stw\,\alWB + \ctw\,\alBW \hsp, &\qquad& \alB = - \ctw\,\alWB + \stw\,\alBW \hsp, 
\nonumber \\
\adW = \stw\,\adWB + \ctw\,\adBW \hsp, &\qquad& \adB = - \ctw\,\adWB + \stw\,\adBW \hsp,  
\nonumber \\
\auW = \stw\,\auWB + \ctw\,\auBW \hsp, &\qquad& \auB = \ctw\,\auWB - \stw\,\auBW \hsp,  
\label{combo}
\end{eqnarray}
where $\alW$, \etc are Wilson coefficients in the Warsaw basis~\cite{Grzadkowski:2010es}. It is worth noting that in SMEFT
we have both triangles and bubbles due to four{-}point vertices like $\PA \phi^0 \PAf \Pf$ \etc. If they are not
included, the anomaly contains an UV{-}divergent term. We introduce
\begin{equation}
\mrC^{\sPA} = - \partial_{\mu}\,\PA_{\mu} \hsp, \;\;
\mrC^{\sPZ} = - \partial_{\mu}\,\PZ_{\mu} + \mrM_{\sPZ}\,\phi^0 \hsp, \;\;
\mrC^{\pm} = - \partial_{\mu}\,\PWpm_{\mu} + \mrM_{\sPW}\,\phi^{\pm} \hsp.
\end{equation}
In the ``diagrammatic'' language of Veltman the validity of the WTST identities is equivalent to the statement that
the $\mrC$ are free fields and any Green's function with one or more external $\mrC\,$-sources is zero.
Considering the effective action $\mrS$, we observe that diagrams determine $\mrS$ only up to an arbitrary choice
of local counterterms and we are free to redefine $\mrS$ by adding to it $\mrS_{\mathrm{ct}}$ with an arbitrary
coefficent~\cite{Preskill:1990fr}. 
\subsubsection{WTST identity for $\mrC^{\sPZ} \PA \PA$} 
In this case we have $\mrC^{\sPZ}(P) \to \PA_{\alpha}(p_1) + \PA_{\beta}(p_2)$. Summing over fermion generations we obtain
\begin{equation}
\Gamma_{\alpha\,\beta} = \frac{g^3}{8\,\pi^2}\,\frac{\stw}{\ctwc}\,g_{_6} 
\,\varepsilon_{\mu\nu\alpha\beta}\,p^{\mu}_1\,p^{\nu}_2\,\sum_{\{\mrg\}}
\lpar 
\frac{m^2_{\Pl}}{\mzs}\,\alWB +
\frac{m^2_{\PQd}}{\mzs}\,\adWB +
2\,\frac{m^2_{\PQu}}{\mzs}\,\auWB \rpar \hsp,
\label{WTSTZAA12}
\end{equation}
for schemes $1$ and $2$. As expected there is no anomaly in $\mrdim = 4$ but there
is one in $\mrdim = 6$ which is mass dependent; 
The standard treatment is that the anomaly can be removed by adding to the Lagrangian a term proportional to
$\varepsilon_{\mu\nu\alpha\beta}\,\phi^0\,\partial^{\mu} \PA^{\alpha}\,\partial^{\nu}\,\PA^{\beta}$.
\subsubsection{WTST identity for $\mrC^{\sPZ} \PZ \PA$} 
In this case we have $\mrC^{\sPZ}(P) \to \PZ_{\alpha}(p_1) + \PA_{\beta}(p_2)$ and obtain
\begin{eqnarray}
\Gamma^{\mrS1}_{\alpha\,\beta} &=& 
\frac{g^3}{32\,\pi^2}\,g_{_6}\,\varepsilon_{\mu\nu\alpha\beta}\,p^{\mu}_1\,p^{\nu}_2\,\sum_{\{\mrg\}}
\Bigl[
2\,\frac{\stw}{\ctwc}\,\bigl( 
\frac{m^2_{\Pl}}{\mzs}\,\alBW +
\frac{m^2_{\PQd}}{\mzs}\,\adBW +
2\,\frac{m^2_{\PQu}}{\mzs}\,\auBW \bigr)
\nonumber \\
{}&+&
\frac{1}{\ctwq}\,\bigl( 
\frac{m^2_{\Pl}}{\mzs}\,\mrv_{\Pl}\,\alWB +
3\,\frac{m^2_{\PQd}}{\mzs}\,\mrv_{\PQd}\,\adWB +
3\,\frac{m^2_{\PQu}}{\mzs}\,\mrv_{\PQu}\,\auWB \bigr)
\nonumber \\
{}&+& \frac{4}{3}\,\frac{\stw}{\ctws}\,\bigl(
3\,\apqt + \apqo - 8\,\apu - 2\,\apd - 3\,\aplt + 3\,\aplo - 6\,\apl \bigr)
\nonumber \\
{}&-& \frac{8}{3}\,\stw\,\bigl(
3\,\apqt + 5\,\apqo - 4\,\apu - \apd - 3\,\aplt + 3\,\aplo - 3\,\apl \bigr)
\Bigr] \hsp,
\label{WTSTZZA1}
\end{eqnarray}
\begin{eqnarray}
\Gamma^{\mrS2}_{\alpha\,\beta} &=& 
\frac{g^3}{64\,\pi^2}\,g_{_6}\,\varepsilon_{\mu\nu\alpha\beta}\,p^{\mu}_1\,p^{\nu}_2\,\sum_{\{\mrg\}}
\Bigl[
4\,\frac{\stw}{\ctwc}\,\bigl( 
\frac{m^2_{\Pl}}{\mzs}\,\alBW +
\frac{m^2_{\PQd}}{\mzs}\,\adBW +
2\,\frac{m^2_{\PQu}}{\mzs}\,\auBW \bigr)
\nonumber \\
{}&+&
\frac{1}{\ctwq}\,\bigl( 
\frac{m^2_{\Pl}}{\mzs}\,\mrv_{\Pl}\,\alWB +
3\,\frac{m^2_{\PQd}}{\mzs}\,\mrv_{\PQd}\,\adWB +
3\,\frac{m^2_{\PQu}}{\mzs}\,\mrv_{\PQu}\,\auWB \bigr)
\nonumber \\
{}&-& 4\,\frac{\stw}{\ctws}\,\bigl(
\apqt + 3\,\apqo + 2\,\apu + \apd - \aplt + \aplo + \apl \bigr)
\Bigr] \hsp,
\label{WTSTZZA2}
\end{eqnarray}
where $\mrv_{\Pf} = 1 - 8\,\mrQ_{\Pf}\,\mrI^3_{\Pf}\,\stws$. There is an anomaly in $\mrdim = 6$ which is mass dependent 
but UV finite and local; however the anomaly is scheme dependent.
\subsubsection{WTST identity for $\mrC^{\sPZ} \PZ \PZ$} 
In this case we have $\mrC^{\sPZ}(P) \to \PZ_{\alpha}(p_1) + \PZ_{\beta}(p_2)$ and obtain
\begin{eqnarray}
\Gamma^{\mrS1}_{\alpha\,\beta} &=& 
\frac{g^3}{16\,\pi^2}\,g_{_6}\,\varepsilon_{\mu\nu\alpha\beta}\,p^{\mu}_1\,p^{\nu}_2\,\sum_{\{\mrg\}}
\Bigl[
\frac{1}{\ctwq}\,\bigl(
\frac{m^2_{\Pl}}{\mzs}\,\mrv_{\Pl}\,\alWB +
3\,\frac{m^2_{\PQd}}{\mzs}\,\mrv_{\PQd}\,\adWB +
3\,\frac{m^2_{\PQu}}{\mzs}\,\mrv_{\PQu}\,\auWB \bigr)
\nonumber \\
{}&+& \frac{8}{3}\,\frac{1}{\ctw}\,\bigl(
3\,\apqt + \apqo - 8\,\apu - 2\,\apd - 3\,\aplt + 3\,\aplo - 6\,\apl \bigr)
\nonumber \\
{}&-& \frac{8}{3}\,\ctw\,\bigl(
3\,\apqt + 5\,\apqo - 4\,\apu - \apd - 3\,\aplt + 3\,\aplo - 3\,\apl \bigr)
\nonumber \\
{}&+& \frac{4}{3}\,\frac{1}{\ctwc}\,\bigl(
- \apqo + 8\,\apu + 2\,\apd - 3\,\aplo + 6\,\apl \bigr)
\Bigr] \hsp,
\label{WTSTZZZ1}
\end{eqnarray}
\begin{eqnarray}
\Gamma^{\mrS2}_{\alpha\,\beta} &=& 
\frac{g^3}{32\,\pi^2}\,g_{_6}\,\varepsilon_{\mu\nu\alpha\beta}\,p^{\mu}_1\,p^{\nu}_2\,\sum_{\{\mrg\}}
\Bigl[
\frac{1}{\ctwq}\,\bigl(
\frac{m^2_{\Pl}}{\mzs}\,\mrv_{\Pl}\,\alBW +
3\,\frac{m^2_{\PQd}}{\mzs}\,\mrv_{\PQd}\,\adBW +
3\,\frac{m^2_{\PQu}}{\mzs}\,\mrv_{\PQu}\,\auBW \bigr)
\nonumber \\
{}&-& \frac{4}{\ctw}\,\bigl(
\apqt + 3\,\apqo + 2\,\apu + \apd - \aplt + \aplo + \apl \bigr)
\nonumber \\
{}&+& \frac{2}{3}\,\frac{1}{\ctwc}\,\bigl(
6\,\apqt + 6\,\apqo + 9\,\apu + 3\,\apd - 6\,\aplt + 2\,\aplo + 5\,\apl \bigr)
\Bigr] \hsp,
\label{WTSTZZZ2}
\end{eqnarray}
where, once again, there is an anomaly in $\mrdim = 6$ which is mass dependent, scheme dependent but UV finite and local.
\subsubsection{WTST identity for $\mrC^{\sPZ} \PWp \PWm$} 
In this case we have $\mrC^{\sPZ}(P) \to \PWm_{\alpha}(p_1) + \PWp_{\beta}(p_2)$ and obtain
\begin{eqnarray}
\Gamma^{\mrS1}_{\alpha\,\beta} &=& 
-\,\frac{g^3}{16\,\pi^2}\,g_{_6}\,\varepsilon_{\mu\nu\alpha\beta}\,p^{\mu}_1\,p^{\nu}_2\,\sum_{\{\mrg\}}
\Bigl[
\bigl(
\frac{m^2_{\Pl}}{\mzs}\,\alWB -
\frac{m^2_{\PQd}}{\mzs}\,\adWB +
\frac{m^2_{\PQu}}{\mzs}\,\auWB \bigr)\,\frac{\stwc}{\ctwc} 
\nonumber \\
{}&+&
\bigl(
\frac{m^2_{\Pl}}{\mzs}\,\alBW -
\frac{m^2_{\PQd}}{\mzs}\,\adBW +
\frac{m^2_{\PQu}}{\mzs}\,\auBW \bigr)\,\frac{\stws}{\ctws} -
4\,\bigl(\apqt - \aplt \bigr)\,\frac{\stws}{\ctw}
\Bigr] \hsp,
\end{eqnarray}
\begin{eqnarray}
\Gamma^{\mrS2}_{\alpha\,\beta} &=& 
-\,\frac{g^3}{24\,\pi^2}\,g_{_6}\,\varepsilon_{\mu\nu\alpha\beta}\,p^{\mu}_1\,p^{\nu}_2\,\sum_{\{\mrg\}}
\Bigl[
\bigl(
\frac{m^2_{\Pl}}{\mzs}\,\alWB -
3\,\frac{m^2_{\PQd}}{\mzs}\,\adWB \bigr)\,\frac{\stw}{\ctwc} 
\nonumber \\
{}&+&
\bigl(
\frac{m^2_{\Pl}}{\mzs}\,\alBW -
3\,\frac{m^2_{\PQd}}{\mzs}\,\adBW \bigr)\,\frac{1}{\ctws} 
\nonumber \\
{}&+&
3\,\bigl(
2\,\frac{m^2_{\PQd}}{\mzs}\,\alWB +
\frac{m^2_{\PQu}}{\mzs}\,\auWB \bigr)\,\frac{\stw}{\ctw} +
3\,\bigl(
2\,\frac{m^2_{\PQd}}{\mzs}\,\alBW +
\frac{m^2_{\PQu}}{\mzs}\,\auBW \bigr) 
\nonumber \\
{}&-&
8\,\bigl(\apqt - \aplt\bigr)\,\frac{\stws}{\ctw}
\Bigr] \hsp.
\end{eqnarray}
\subsubsection{WTST identity for $\mrC^{\sPA} \PWp \PWm$} 
In this case we have $\mrC^{\sPA}(P) \to \PWm_{\alpha}(p_1) + \PWp_{\beta}(p_2)$ and obtain
\begin{eqnarray}
\Gamma^{\mrS1}_{\alpha\,\beta} &=& 
\frac{g^3}{16\,\pi^2}\,g_{_6}\,\varepsilon_{\mu\nu\alpha\beta}\,p^{\mu}_1\,p^{\nu}_2\,\sum_{\{\mrg\}}
\Bigl[
\bigl(
\frac{m^2_{\Pl}}{\mzs}\,\alWB -
\frac{m^2_{\PQd}}{\mzs}\,\adWB +
\frac{m^2_{\PQu}}{\mzs}\,\auWB \bigr)\,\frac{\stws}{\ctws} 
\nonumber \\
{}&+&
\bigl(
\frac{m^2_{\Pl}}{\mzs}\,\alBW -
\frac{m^2_{\PQd}}{\mzs}\,\adBW +
\frac{m^2_{\PQu}}{\mzs}\,\auBW \bigr)\,\frac{\stw}{\ctw} -
4\,\bigl(\apqt - \aplt \bigr)\,\stw
\Bigr] \hsp,
\end{eqnarray}
\begin{eqnarray}
\Gamma^{\mrS2}_{\alpha\,\beta} &=& 
-\,\frac{g^3}{8\,\pi^2}\,g_{_6}\,\varepsilon_{\mu\nu\alpha\beta}\,p^{\mu}_1\,p^{\nu}_2\,\sum_{\{\mrg\}}
\Bigl[
\bigl(
2\,\frac{m^2_{\PQd}}{\mzs}\,\adWB +
\frac{m^2_{\PQu}}{\mzs}\,\auWB \bigr)\,\frac{\stws}{\ctws} 
\nonumber \\
{}&+&
\bigl(
2\,\frac{m^2_{\PQd}}{\mzs}\,\adBW +
\frac{m^2_{\PQu}}{\mzs}\,\auBW \bigr)\,\frac{\stw}{\ctw} +
\frac{8}{3}\,\bigl(\apqt - \aplt \bigr)\,\stw
\Bigr] \hsp,
\end{eqnarray}
In this case the anomaly can be removed by adding to the Lagrangian a term proportional to
$\varepsilon^{\mu\alpha\beta\nu}\,\mrA_{\nu}\,\bigl(\mrW^+_{\beta} \partial_{\mu} \mrW^-_{\alpha} - 
\mrW^-_{\alpha} \partial_{\mu} \mrW^+_{\beta} \bigr)$.
\subsubsection{WTST identity for $\mrC^{\sPA} \PZ \PZ$ and $\mrC^{\sPA}\PZ\PA$} 
We only present the results for scheme $1$. They are
\begin{eqnarray}
\Gamma^{\mrS1}_{\alpha\,\beta} &=& 
\frac{g^3}{12\,\pi^2}\,g_{_6}\,\varepsilon_{\mu\nu\alpha\beta}\,p^{\mu}_1\,p^{\nu}_2\,\sum_{\{\mrg\}}
\Bigl[
\frac{3}{2}\,\Bigl(
\frac{m^2_{\Pl}}{\mzs}\,\alBW + \frac{m^2_{\PQd}}{\mzs}\,\adBW + 2\,\frac{m^2_{\PQu}}{\mzs}\,\auBW \Bigr)\,\frac{\stw}{\ctwc}
\nonumber \\
&+&
\bigl(
3\,\apqt - 3\,\aplt - 8\,\apu - 2\,\apd + \apqo + 3\,\aplo - 6\,\apl \bigr)\,\frac{\stw}{\ctws} 
\nonumber \\
&-& 2\,\bigl(
3\,\apqt - 3\,\aplt - 4\,\apu - \apd + 5\,\apqo + 3\,\aplo - 3\,\apl \bigr)\,\stw
\Bigr] \hsp, 
\end{eqnarray}
for $\mrC^{\sPA} \PZ \PZ$ and
\begin{eqnarray}
\Gamma^{\mrS1}_{\alpha\,\beta} &=& 
\frac{g^3}{16\,\pi^2}\,g_{_6}\,\varepsilon_{\mu\nu\alpha\beta}\,p^{\mu}_1\,p^{\nu}_2\,\sum_{\{\mrg\}}
\Bigl[
\Bigl(
\frac{m^2_{\Pl}}{\mzs}\,\alBW + \frac{m^2_{\PQd}}{\mzs}\,\adBW + 2\,\frac{m^2_{\PQu}}{\mzs}\,\auBW \Bigr)\,\frac{\stw}{\ctwc}
\nonumber \\
&-& \frac{4}{3}\,
\bigl(
3\,\apqt - 3\,\aplt - 4\,\apu - \apd + 5\,\apqo + 3\,\aplo - 3\,\apl \bigr)\,\frac{\stws}{\ctw} 
\Bigr] \hsp,
\end{eqnarray}
for $\mrC^{\sPA} \PZ \PA$. 
\subsubsection{WTST identity for $\mrC^{\sPZ} \PG \PG$} 
In this case (where $\PG^a_{\mu}$ is the gluon field) we have $\mrC^{\sPZ}(P) \to \PG^a_{\alpha}(p_1) + \PG^b_{\beta}(p_2)$ 
and obtain
\begin{equation}
\Gamma^{a\,b}_{\alpha\,\beta} = 
\frac{g g^2_{\mrS}}{16\,\pi^2\,\ctws}\,g_{_6}\,\varepsilon_{\mu\nu\alpha\beta}\,p^{\mu}_1\,p^{\nu}_2\,
\delta^{a\,b}\,\sum_{\{\mrg\}}\,\bigl(
\frac{m^2_{\PQd}}{\mzs}\,\adG +
\frac{m^2_{\PQu}}{\mzs}\,\auG \bigr) \hsp,
\end{equation}
with a sum over the quark generations and where $g_{\mrS}$ is the $SU(3)$ coupling constant. There is an anomaly 
in $\mrdim = 6$ which is mass dependent but UV finite and local.
\subsubsection{WTST identity for $\mrC^{-} \mrW^+ \PZ$ and $\mrC^{-}\mrW^+\PA$} 
In these two cases we have found no anomaly in scheme $1$.
However, a technical remark is needed: for amplitudes having a Born term we must take into account the relation between
bare and renormalized parameters and also include Dyson{-}resummation of the propagators (when needed). The identity
reads as follows:
\begin{equation}
i\,p^{\lambda}\,\mrA^{\sPW}_{\lambda \alpha \beta} + \mrM\,\mrA^{\phi}_{\alpha\beta} = 0 \hsp,
\end{equation}
where $\mrM$ is the bare $\mrW$ mass and we use $\mrM_0 = \mrM/\ctw$ for the bare $\mrZ$ mass. Inside and in front of
loops we will use the on{-}shell masses, $\mrM_{\sPW}$ and $\mrM_{\sPZ}$. However, in this case there is a lowest
order where the relation between bare and on{-}shell masses must be corrected at $\mcO(g^2)$, involving the
corresponding self{-}energies.

There is an important remark to be made here: consider the SM ($\mrdim = 4$), where there is no anomaly; consider
now the SM with one electron and one neutrino, scheme $1$ gives zero anomaly which is in conflict with the
master formula for the one{-}loop anomalies given in \Bref{Sterling:1981za}. The anomalies require a counterterm
proportional to
\begin{equation}
\varepsilon_{\mu\nu\alpha\beta}\,\Bigl( \mrF_{\mu\nu}\,\mrG^a_{\alpha\beta}\,\phi^a + \mrF_{\mu\nu}\,\mrF_{\alpha\beta}\,\phi^0\,
\tan\theta \Bigr) \hsp,
\end{equation}      
where $\mrF_{\mu\nu}$ is the field strength of the original $U(1)$ vector boson, $\mrG^a_{\mu\nu}$ is the field
strength of the original $SU(2)$ vector boson and $\phi^a$ is the Higgs{-}Kibble ghost (with $\phi^3 \equiv \phi^0$).
The correct result is reproduced with scheme $2$; here the $\mrC^-(p_1) \mrZ_{\mu}(P) \mrW^+_{\beta}(p_2)$ identity 
in SMEFT is given by
{\footnotesize{
\begin{eqnarray}
\Gamma^{\mrS2}_{\beta\,\mu} &=& 
\frac{g^3}{192\,\pi^2\,\mrM^2_{\sPW}}\,g_{_6}\,\varepsilon_{\alpha\nu\beta\mu}\,p^{\alpha}_1\,p^{\nu}_2\,
\nonumber \\
&\times& \sum_{\{\mrg\}}
\Bigl[
\Bigl( 5\,m^2_{\Pl}\,\alWB + 15\,m^2_{\PQd}\,\adWB - 3\,m_{\PQu} m_{\PQd}\,(\auWB + \adWB) + 
15\,m^2_{\PQu}\,\auWB \Bigr)\,\frac{\stw}{\ctw}
\nonumber \\
&+& 12\,\Bigl( m^2_{\Pl}\,\alBW + 3\,m^2_{\PQd}\,\adBW + 3\,m^2_{\PQu}\,\auBW \Bigr)
\nonumber \\
&-& 2\,\Bigl( 7\,m^2_{\Pl}\,\alWB + 5\,m^2_{\PQd}\,\adWB + m_{\PQu} m_{\PQd}\,(\auWB - \adWB) +
13\,m^2_{\PQu}\,\auWB \Bigr)\,\frac{\stwc}{\ctw}
\nonumber \\
&-& 2\,\Bigl( 7\,m^2_{\Pl}\,\alBW + 5\,m^2_{\PQd}\,\adBW + m_{\PQu} m_{\PQd}\,(\auBW - \adBW) +
13\,m^2_{\PQu}\,\auBW \Bigr)\,\stws
\nonumber \\
&+& 32\,\mrM^2_{\sPW}\Bigl( \apqt - \aplt \Bigr)\,\frac{\stws}{\ctw} -
32\,\mrM^2_{\sPW}\,\Bigl( 3\,\apqo + \aplo \Bigr)\,\frac{1}{\ctw} \Bigr] \hsp.
\end{eqnarray}
}}
\subsection{SMEFT and anomalies: conclusions} 
There are SMEFT anomalies, UV{-}finite, local, mass{-}dependent and scheme{-}dependent. There are different options, 
for instance use consistently a scheme, \myeg the (naive) anti{-}commuting $\gamma^5$ scheme or
the Veltman scheme and introduce counterterms. 
This procedure is the one which we could use working with the SM with one electron and one neutrino: due to the anomaly, 
the WTST indentities break down at the one{-}loop level. We then introduce counterterms and the identities are restored 
but the terms are of a non{-}renormalizable nature and they give rise to infinities, at the earliest at the two{-}loop 
level, as described in \Bref{Sterling:1981za}. 

Alternatively, we could cancel the anomalies. For instance, within the anti{-}commuting $\gamma^5$ scheme 
we obtain the following relations:
\begin{eqnarray}
{}&{}& \afB = \afW = 0 \quad \forall \mrf \hsp, \quad \auG = \adG = 0 \hsp,
\nonumber \\ 
{}&{}& \aplt = \apqt \hsp, \; \aplo = - 3\,\apqo \hsp, \; \apl = - \frac{1}{3}\,\Bigl(
4\,\apqo + 4\,\apu + \apd \Bigr) \hsp.
\label{relations}
\end{eqnarray}
Similar relations, obtained within the Veltman scheme, are less ``transparent''. It is worth noting that in the SMEFT 
there is a breakdown of the WTST identities, no matter which scheme is used. 
As long as the anomalies are UV{-}finite and local (which is the ``tested'' case up to three{-}point functions)
they can be removed, within a given scheme, by counterterms.

One possible question is whether to include \eqn{relations} in fits~\cite{Minn:1986ba,Feruglio:2020kfq}. We have shown that 
the cancellation of SMEFT anomalies is scheme dependent (extending to the full SMEFT the work of \Bref{Feruglio:2020kfq})
but there is also another point~\cite{Bonnefoy:2020tyv}: what 
happens if we have a UV complete and anomaly{-}free underlying theory whose low{-}energy behavior violates the identities? 
There doesn't seem to be any obvious solution if we insist with this strategy; note that only
$\afB$ and $\afW$ are loop{-}generated~\cite{Einhorn:2013kja} $\mrdim = 6$ operators that we have inserted in loops and
mass{-}dependent anomalies are due to loop{-}generated operators. Having said that, the fact that using the relations
in \eqn{relations} is enough to cancel the $\mrC^{\sPZ}$ and $\mrC^{\sPA}$ anomalies remains an intriguing consideration 
involving the fermions.

A final comment follows from the work of \Bref{Sterling:1981za}: 
one should investigate the divergences arising in higher orders from these counterterms, something happenning
if the counterterms are part of a loop (these diagrams must be seen as two{-}loop diagrams). 
The result is that there are several logarithmically divergent amplitudes, \myeg 
an anomalous magnetic moment for the fermions. 

WTST identities for four{-}point functions show new interesting aspect to be presented elsewhere.
\section{Conclusions}
It is straightforward to see that when it comes to confronting theoretical predictions with experimental results
there will always be a reference to a work of Martinus Veltman.


\bibliography{Passarino_4_Veltman}
\bibliographystyle{JHEP}

%
%
\end{document}